# Coherently driven photonic de Broglie Sagnac interferometer

Byoung S. Ham

School of Electrical Engineering and Computer Science, Gwangju Institute of Science and Technology, 123 Chumdangwagi-ro, Buk-gu, Gwangju 61005, S. Korea

(Submitted on February 18, 2020)

Quantum measurements have been intensively researched over decades due to quantum advantage of Heisenberg limit beating the standard quantum limit toward potential applications of quantum metrology. The kernel of quantum measurements is in the quantum correlation between bipartite photon pairs or squeezed light quenched by one parameter over corresponding noncommuting variable satisfying Heisenberg uncertainty principle. As a result, quantum measurements bring a quantum gain of the square root law in measurement sensitivity. Photonic de Broglie waves (PBW) have been the key feature of such a gain in quantum metrology especially for phase resolution enhancement beyond the classical limit of Rayleigh criterion or simply the diffraction limit. Due to extremely low efficiency of higher-order entangled photon pair generations such as a NOON state, however, the implementation of PBW for quantum metrology has been severely limited. Here, a completely different mechanism of quantum measurements is introduced for a new type of PBW and presented for its potential application of a modified Sagnac interferometer, where the resolution enhancement is several orders of magnitude higher than its classical counterpart.

1. **INTRODUCTION**

Measurement is a physical process for a physical quantity such as intensity of electric fields. The accuracy of measurement is denoted by statistical errors where the error can be reduced with trial number N due to the classical low of $1/\sqrt{N}$ [1]. Thus, the signal to noise ratio increases as the light is brighter until saturated. This is the fundamental law of classical physics governed by Poisson statistics, where the classicality represents for independence among trials or photons [2]. If there is quantum correlation among the trials or photons, however, the statistical error can be reduced more proportional to $1/N$, where the square root enhancement in the measurement sensitivity or phase resolution is due to the quantum gain originating in the quantum correlation which cannot be obtained by a classical means [3-5]. Such a quantum correlation is represented by a nonclassical feature demonstrated in entangled photon pairs [6-8], squeezed light [9], and a Fock state [10]. The direct proof of this measurement gain with quantum correlation is in the photonic de Broglie waves (PBW), where the phase resolution is enhanced by N of entangled photons such as in a NOON state and a Schrödinger's cat represented by $(|N\rangle_A|0\rangle_B + |0\rangle_A|N\rangle_B)/\sqrt{2}$ [11-13]. This also results in the Heisenberg limit beating the classical standard quantum limit, proving that the classical limit can be overcome by properly selecting measurement methods [1-17]. The practical difficulty of PBW is, however, due to the indeterminacy of NOON state generation. Thus, the implementation of quantum metrology for such as frequency standard [14], imaging [15], spectroscopy [16], and lithography [17] have been severely limited.

Since the first demonstration in 1913 [18], the Sagnac interferometer (SI) has been implemented for optical [19] and matter-wave [20] interferometry as well as atomic spectroscopy [21] and gravitational wave detection [22]. Due to the limited Sagnac effect [23], however, the SI cannot be applied for stand-alone inertial navigation systems of aircrafts, rockets, submarines, and space vehicles. Here, a completely different mechanism of PBW, the so-called coherence PBW, is introduced and presented for the quantum advantage in SI with a few orders of magnitude enhancement in phase resolution. For the introduction of coherence PBW, a cross-coupled double (CCD) Mach-Zehnder interferometer (MZI) is investigated for the coherence control of phase in each MZI, resulting in the nonclassical feature of PBW. Unlike conventional PBW [11-13], the physics of coherence PBW is in the quantum superposition control for an asymmetrically coupled double MZI [24]. Using the coherence PBW, a photonic de Broglie Sagnac interferometer (PBSI) is presented for the quantum advantage of enhanced phase resolution in a few orders of magnitude higher than the classical limit. Owing to its on-demand control, the present method of coherence PBW opens a door to coherence-quantum metrology such as quantum lithography and sensing as well as quantum inertial navigation and geodesy in a pure coherence manner for an optical regime.

The fundamental limit of phase resolution in classical physics such as SI is given by the Rayleigh criterion governed by the wavelength $\lambda_0$ of light. Using PBW, however, the classical limit can be overcome owing to the nonclassical feature of entanglement or squeezing, whose photonic de Broglie wavelength $\lambda_B$ is shortened by the degree of nonclassicality or N in a NOON state [11-13]: $\lambda_B = \lambda_0/N$. For example, if N=2, the degree of nonclassicality is doubled compared with N=1 (classical limit). Thus, the phase resolution in PBW is enhanced by N. Although this N results in the quantum gain of $\sqrt{N}$ in sensitivity or measurements, our interest is in the linear enhancement over the diffraction limit with N in $\lambda_B$. The degree of nonclassicality in the NOON state is represented by how many correlated photons (N) are involved in measurements. The resulting quantum enhancement in measurement sensitivity using high N PBW has been well demonstrated in a single MZI [2,7,11-13,17]. Because achieving high N is extremely difficult, however, the implementation of quantum metrology is also challenging. On the contrary, as will be analyzed in Fig. 1, the degree of N here in coherence PBW is somewhat extraordinary, where only coherence control of phases in the coupled MZI matters. With the present coherence PBW, therefore, such a limiting factor not only in classical physics but also in conventional PBW is completely overcome, and the quantum advantage is directly applied for a quantum metrology based on PBSI. This is the unprecedented discovery in both classical and quantum physics.

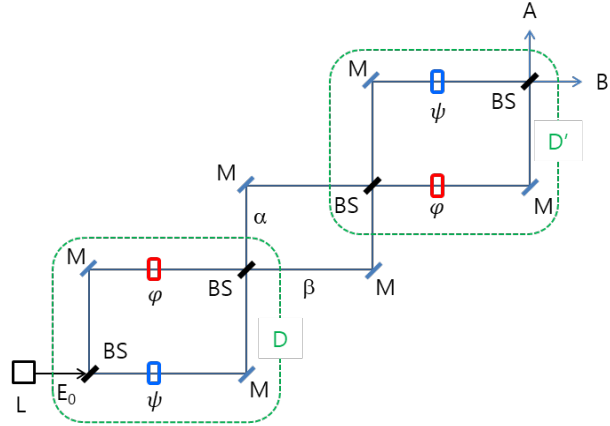

Fig. 1. A cross-coupled double-MZI for photonic de Broglie wavelength generation.
L: laser, M: mirror, BS: beam splitter, $\varphi/\psi$: phase controller. $E_0$ is a coherent light.

## 2. PHOTONIC DE BROGLIE SAGNAC

Figure 1 shows a CCD-MZI as a basic building block of coherence PBW. Unlike a conventional quantum measurement scheme using nonclassical lights in two ports [2-17], there is only one input of classical light from a typical laser. It should be noted that the coherent light $E_0$ in Fig. 1 is not a few photons but a phase coherent bright light. As proved already, the single input MZI satisfies the same quantum characteristics of anticorrelation, the so-called a HOM dip, through a beam splitter (BS) [25]. In other words, each MZI block in Fig. 1 represents for a quantum device resulting in anticorrelation. As already investigated on a BS [26], the basic physics of coherence PBW is originated in quantum superposition between two paths of each MZI via coherence control of asymmetrically connected phase shifter $\varphi$ (as well as $\psi$), resulting in double quantum superposition in the coupled MZI system of Fig. 1 [24]. Here, the anticorrelation on a BS (or MZI) represents for maximal coherence of perfect indistinguishability between two inputs (or paths) [27], where the phase relation must be satisfied by $\phi_{in} = \frac{(2m+1)\pi}{2}$ $(m = 0,1,2,...)$ on a BS [26]. Such quantum correlation of entanglement has already been demonstrated by applying such a phase constraint to two independently trapped atoms [28].

In Fig. 1, the phase control has been modified for the Sagnac effect in SI, where two phases $\varphi$ and $\psi$ must satisfy the anti-phase relation $(\psi = -\varphi)$, where counterpropagaitng lights induce an opposite phase equivalent to the antiphase relation with a relativistic time (phase) delay $\Delta t$: $\Delta t = \frac{4A\Omega}{c^2}$; A is the area of SI's closed loop; $\Omega$ is the rotation rate. Under this antiphase condition, the output fields α and β in the first block D results in a nonclassical feature of anticorrelation if $\Delta\phi$ $(2\varphi) = \pm m\pi$ (m = 0,1,2, … ). Because the Sagnac effect cannot satisfy such as big phase shift $(2\varphi = \frac{8\pi A\Omega}{c\lambda_B})$, PBSI may be preset on this antiphase condition and the Sagnac effect may be monitored. The output field is of course unidirectional either into α or β depending on m. Although the input light $E_0$ is a coherent source, the output field (α and β) can be nonclassical via anticorrelation [26] and enters the second block D', resulting in the output A and B through the same mechanism as in D. The asymmetric configuration of phase control in Fig. 1 is for ordered quantum superposition in the coupled MZI [24]. This is the heart of coherence PBW, resulting in the quantum gain in measurements or phase resolution. As proved, PBW is an inherent quantum nature and cannot be obtained classically [11-13]. So does the coherence PBW if the same phenomenon is shown (see Figs. 2 and 3).

The followings are matrix representations for analytic solutions in the first MZI (block D) of Fig. 1 as functions of the independent phases $\varphi$ and $\psi$:

$$\begin{bmatrix}\alpha\\\beta\end{bmatrix} = [BS][\Theta][BS]\begin{bmatrix}E_0\\0\end{bmatrix},$$
$$= \frac{1}{2}e^{i\psi}\begin{bmatrix}(1-e^{i(\varphi-\psi)}) & i(1+e^{i(\varphi-\psi)})\\ i(1+e^{i(\varphi-\psi)}) & -(1-e^{i(\varphi-\psi)})\end{bmatrix}\begin{bmatrix}E_0\\0\end{bmatrix}, \quad (1)$$

where $[BS]$ and $[\Theta]$ are $\frac{1}{\sqrt{2}}\begin{bmatrix}1 & i\\ i & 1\end{bmatrix}$ and $\begin{bmatrix}e^{i\psi} & 0\\ 0 & e^{i\varphi}\end{bmatrix}$, respectively (see the Supplementary Information). Thus, the light intensity of α and β become, respectively:

$I_\alpha = \frac{I_0}{2}[1 - \cos(\varphi - \psi)]$ $(= I_0 \sin^2[(\varphi - \psi)/2])$, (2-1)

$I_\beta = \frac{I_0}{2}[1 + \cos(\varphi - \psi)]$ $(= I_0 \cos^2[(\varphi - \psi)/2])$, (2-2)

where $I_0$ is the intensity of $E_0$. Compared with the one-phase–based MZI [24], whose modulation period is $2\pi$, equations (2-1) and (2-2) show a half-modulation period if $\psi = -\varphi$ (see Fig. 2(b)).

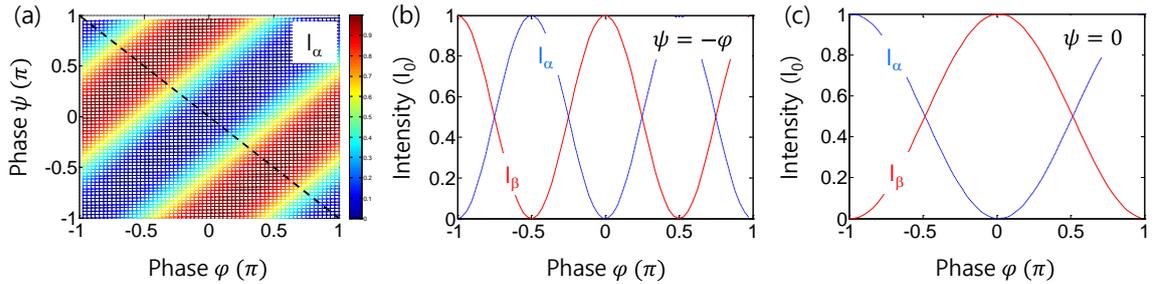

Fig. 2. Numerical calculations for the first MZI outputs in Fig. 1. (a) $I_\alpha$ of equation (2-1). Intensity for (b) $\psi = -\varphi$ and (c) $\psi = 0$. For $I_\beta$ of equation (2-1), see the Supplementary Information.

In Fig. 2, the first MZI outputs $I_\alpha$ and $I_\beta$ are numerically calculated using equations (2-1) and (2-2). As shown in Fig. 2(a), the intensity modulation period depends on the relative phase between $\varphi$ and $\psi$, where it is maximized if the antiphase $(\psi = -\varphi)$ relation is satisfied (see the dashed line). In Fig. 2(b), the anticorrelation $g^{(2)} = 0$ is shown at $\varphi = \pm m\pi/2$, where the $\pi/2$ modulation period (not shown) is the same as the quantum PBW with N=4. Figure 2(c) shows a reference of a classical limit as in a typical MZI if $\psi = 0$, where it represents for the Rayleigh criterion or the diffraction limit of $\lambda_0/2$. According to the classical physics, the phase/spatial resolution is strongly limited by the wavelength $\lambda_0$ of light. Thus, the condition of antiphase in Fig. 1 results in $\lambda_0/4$ ($\pi/2$) in $g^{(2)}$ as a limit of SI (see Fig. 2(b)). As mentioned above, the antiphase relation is automatically fulfilled by the Sagnac effect in a rotating SI: $\Delta\phi = \phi_{ccw} - \phi_{vw} = 2\varphi$ (see also Figs. 2(b) and 3(b)) [18-23,29-35].

In the CCD-MZI of Fig. 1, the matrix representation for the final outputs A and B is as follows (see the Supplementary Information):

$$\begin{bmatrix} A \\ B \end{bmatrix} = [BS][\Theta'][BS]\begin{bmatrix} \alpha \\ \beta \end{bmatrix},$$
$$= e^{i(\varphi+\psi)} \begin{bmatrix} \cos(\varphi-\psi) & \sin(\varphi-\psi) \\ -\sin(\varphi-\psi) & \cos(\varphi-\psi) \end{bmatrix} \begin{bmatrix} E_0 \\ 0 \end{bmatrix}, \quad (3)$$

where $[\Theta'] = \begin{bmatrix} e^{i\varphi} & 0 \\ 0 & e^{i\psi} \end{bmatrix}$. Thus, intensities $I_A$ and $I_B$ of A and B are, respectively:

$$I_A = I_0 \cos^2(\varphi-\psi), \quad (4\text{-}1)$$
$$I_B = I_0 \sin^2(\varphi-\psi). \quad (4\text{-}2)$$

As numerically demonstrated in Fig. 3(a), the intensity modulation frequency of $I_A$ with respect to $\varphi$ is twice higher than Fig. 2(a). The maximum modulation frequency appears when the antiphase condition is met as shown in Fig. 3(b). The lowest modulation frequency (or resolution) is achieved at $\psi = 0$ as shown in Fig. 3(c), which is equivalent to Fig. 2(b) for the case of four-photon quantum PBW [12]. A a result, the coherence PBW in Fig. 3(b) is equivalent to the case of N=8 in quantum PBW [13] and demonstrates for the nonclassical feature of PBW beyond the classical limit. From Figs. 2 and 3, it is not difficult to say that an n-folded modulation frequency in $\lambda_B$ can be obtained in a recursive configuration of CCD-MZI in Fig. 1 (discussed in Fig. 4).

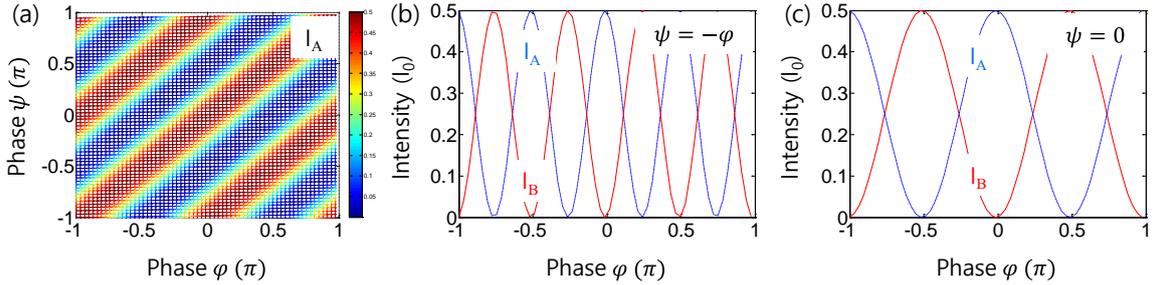

Fig. 3. Numerical calculations for the second MZI outputs in Fig. 1. (a) $I_A$ for equation (4-1). Output intensity for (b) $\psi = -\varphi$ and (c) $\psi = 0$. For $I_B$, see the Supplementary Information.

The application of the discrete control of $\lambda_B$ in Figs. 2 and 3 has already been introduced in a series of a single-phase coupled MZI scheme ($\psi = 0$), resulting in the intensity modulation frequency proportional to $cos(2n\varphi)$ for the output A and $sin(2n\varphi)$ for the output B, where n is the repetition number of the single-phase coupled MZI [24]. Compared with conventional quantum PBW with $\lambda_B = \lambda_0/2N$ in the intensity correlation $g^{(2)}$ [11-13], the coherence PBW shows a corresponding relation of $\lambda_B = \lambda_0/4n$, where the additional factor of 2 comes from the double MZI scheme. As the indistinguishability (maximum coherence) is the prerequisite for the entangled photon pairs due to undecided phase between two input photons in quantum PBW [11-13], the same relation occurs in the coherence PBW for the path superposition in each MZI for the shared input field satisfying anticorrelation. Thus, the coherence PBW is equivalent to quantum PBW, where a specific antiphase relation in MZI plays a key role for the generatioin of nonclassical feature of $\lambda_B$.

For the recursive configuration using CCD-MZI of Fig. 1, a cavity PBSI of Fig. 4(a) is introduced for the SI application to quantum metrology. Unlike individual PBWs discussed in ref. 24, Fig. 4(a) induces a collective version of coherence PBWs in an optical cavity. Thus, each output A or B is resulted from many-wave interference among different $\lambda_B$s of coherence PBW. The cavity PBSI of Fig. 4(a) becomes an intrinsic quantum device satisfying anticorrelation if the antiphase condition is met for the MZI path superposition of all ordered coherence

PBWs. Here, the minimum (effective) $\lambda_B$ is determined by the optical *Finesse* $\mathcal{F}$: $\mathcal{F} = \frac{\pi r}{1-r^2}$. Due to the reflection (transmission) coefficient $r$ (t) on the cavity mirror C in Fig. 4(a), the ordered amplitudes of PBW are gradually reduced as the order n increases, where n is the repetition number of CCD-MZI. The related numerical calculations are shown in Figs. 4(b)~(h).

## 3. ANALYSIS

Starting from equation (3), the resultant $n^{th}$ order of PBW in the cavity PBSI of Fig. 4(a) is obtained as follows (see the Supplementary Information):

$$(E_A)^n = (-1)^n T r^{(n-1)} E_0 \cos(2n\varphi), \qquad (5\text{-}1)$$

$$(E_B)^n = (-1)^{n+1} T r^{(n-1)} E_0 \sin(2n\varphi), \qquad (5\text{-}2)$$

where $T = t^2$. Here, the phase shift accumulated on each round-trip light in the cavity is assumed to be either $\pi$ or $2\pi$, and all optics inside the cavity are lossless to the light. If the phase shift is $2\pi$, i.e., $2n\varphi = \pm 2m\pi$, each even ordered field is perfectly cancelled out by each odd ordered one due to the prefactor of $(-1)^n$ or $(-1)^{n+1}$. Thus, the sum of all n-ordered amplitudes in each field of equations (5-1) and (5-2) becomes zero at $\varphi \to \varphi_{mn}^d = \pm m\pi/n$, where n=1,2,3… If $n \gg 1$, higher order components ($n \geq 2$) of $(E_A)^n$ locate nearly everywhere regularly in the phase axis due to $n^{-1}$ factor in $\varphi_{mn}^d$, and each $(E_A)^{n+1}$ has a sign flip with respect to each $(E_A)^n$, resulting in a complete destructive interference for all n, except for $\varphi = \pm m\pi/2$ (discussed below).

On the contrary, if the phase shift is $\pm\pi$ during a round trip in the cavity, i.e., $2n\varphi = \pm(2m+1)\pi$, equations (5-1) and (5-2) result in a constructive interference due to the $\pi$–phase shift-caused sign flip between the $n^{th}$ and $(n+1)^{th}$ components. This sign flip exactly compensates for the prefactor-induced sign flip (see the arrows in Fig. 4(b)). Thus, all components of $(E_A)^n$ interfere constructively at $\varphi_{mn} = \pm\left(\frac{(2m+1)\pi}{2}\right)$.

For a brief review of the constructive interference in PBSI, let's set m=0 and discuss equation (5-1) with $\varphi_{0n}^c = \pm\left(\frac{\pi}{2}\right)\frac{1}{n}$. For the first order n=1, $(E_A)^{n=1} = -(-1)^1 T r^0 E_0 = TE_0$ at $\varphi_{01}^c = \pm\frac{\pi}{2}$ (see the blue curve in Fig. 4(b)). Remember that the CCD-MZI has a $\frac{\pi}{2}$ modulation period in intensity in Fig. 2, resulting in a $\pi$ modulation in amplitude. Thus, there is a sign flip whenever the phase is even multiplied. If the phase is odd multiplied, there is no sign flip in $(E_A)^n$. For the second order n=2, $(E_A)^{n=2} = -(-1)^2 T r^1 E_0 = -TrE_0$ at $\varphi_{02}^c = \pm\frac{\pi}{4}$. Thus, $(E_A)^{n=2}$ becomes flipped over to $TrE_0$ at $\varphi = \pm\frac{\pi}{2}$ due to the even multiple in the phase, i.e., $\varphi = 2\varphi_{12}^c$. For the third order, n=3, $(E_A)^{n=3} = -(-1)^3 Tr^2 E_0 = Tr^2 E_0$, at $\varphi_{13} = \pm\frac{\pi}{6}$. Thus, $(E_A)^{n=3}$ has no sign flip at $\varphi = \pm\frac{\pi}{2}$ due to an odd multiple, resulting in $Tr^2 E_0$ at $\varphi = 3\varphi_{13}^c$ (see the green curve in Fig. 4(b)). For the fourth order, n=4, $(E_A)^{n=4} = -(-1)^4 Tr^3 E_0 = -Tr^3 E_0$ is satisfied at $\varphi_{13} = \pm\frac{\pi}{8}$. Thus, $(E_A)^{n=4}$ is flipped over and becomes $Tr^3 E_0$ at $\varphi = \pm\frac{\pi}{2}$ due to the even multiple, i.e., $\varphi = 4\varphi_{13}^c$. As a result, all n-ordered components in equation (5-1) constructively interfere at $\varphi_{mn}^c = \pm\left(\frac{(2m+1)\pi}{2}\right)$ due to the resultant in-phase relation among $(E_A)^n$.

For the infinite series of $(E_A)^n$ and $(E_B)^n$ in equations (5-1) and (5-2) at $\varphi = \pm\left(\frac{(2m+1)\pi}{2}\right)$, a general solution for the amplitude sum $E_A$ and $E_B$ is obtained analytically as follows (see the Supplementary Information):

$E_A = TE_0 \sum_{n=1} r^{(n-1)}$, (6-1)
$E_B = 0$, (6-2)

where the prefactor $(-1)^n$ is cancelled out by the accumulated phase $e^{in\delta}$ in the n-round trip(s), where $\delta = \pi$ (see Figs. 4(c) and (d)). Using Taylor expansion, the amplitude sum becomes $E_A = TE_0 \frac{1}{1-r} = E_0(1+r)$. Thus, the finial output intensity along the port A is $I_A = I_0(1+r)^2$, where $I_A = E_A E_A^*$ and $I_0 = |E_0|^2$. For a high reflectance cavity mirror, i.e., $r\sim 1$, the upper bound of the output intensity $I_A$ becomes a quadruple of the input intensity $I_0$. For a nearly transparent cavity mirror ($r\sim 0$), the output intensity $I_A$ shows its lower bound at $I_0$.

In other words, $I_0 \leq I_A \leq 4I_0$ is satisfied in PBSI at $\varphi = \pm\left(\frac{(2m+1)\pi}{2}\right)$, otherwise $I_A = 0$ (see Figs. 4(e)~(g)). Due to the extremely law duty cycle in $I_A$, the maximum $I_A$ does not violate the energy conservation law.

For equation $E_B$, it is zero due to $sin(2n\varphi) = 0$ at $\varphi = \pm\left(\frac{(2m+1)\pi}{2}\right)$. However, there are nonzero sidebands in $E_B$ with intensity maxima of $I_0$ (see Figs. 4(f) and (h)). Details are discussed in section 4 (NUMERICDAL CALCULATIONS).

## 4. NUMERICAL CALCULATIONS

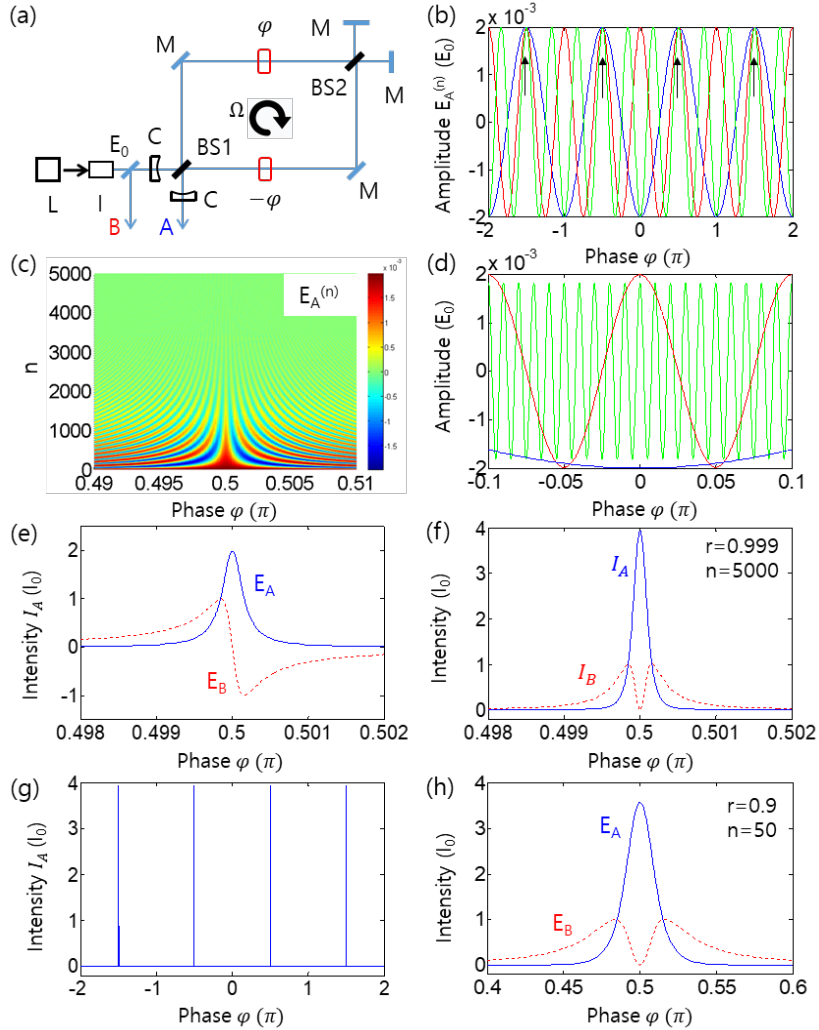

Figure 4. A Cavity photonic de Broglie Sagnac interferometer and its numerical calculations. (a) A schematic of cavity PBSI based on the present coherence PBW. (b)-(f) Numerical calculations for (a). (b) Each ordered field ($E_A$) for the output A in (a): Blue, n=1; Red, n=2; Green, n=3; n represents number of light circulation in (a). The arrows indicate common phase bases resulting in constructive interference. (c) All ordered components of $E_A$. (g) Individual $(E_A)^n$. Blue: n=1; Red: n=10; Green: n=100. (e) Details of amplitude sum for (c): $E_B$ is at the output B. (f) Details of output intensity of $I_A$ for (c). (g) Output intensity of $I_A$ for a broad range of phase. (h) Output intensity for the case of r=0.9 and n=50. L: laser, I: optical isolator, C: cavity mirror, M: mirror, BS: 50/50 beam splitter. For numerical calculations, n=5,000 and r=0.999 are set for (a)~(g).

Figures 4(b)~4(g) show numerical calculations for PBSI by solving equations (5-1) and (5-2) for n=5,000 and r=0.999. As analyzed above, $(E_A)^n$ results in constructive interference at $\varphi = \pm\left(\frac{(2m+1)\pi}{2}\right)$: see the Supplementary Information for equation (5-2). The origin of constructive interference is the π phase gain in each round trip, resulting in the sign flip between n and n+1 (see Fig. 4(b)). The effective n is of course determined by *Finesse* (or *r*) as shown in Fig. 4(c) (see the Supplementary Information). Figure 4(d) shows some examples of coherence PBW generated in the cavity PBSI. As a result, all different $\lambda_B$ interfere together, resulting in the constructive interference at a specific phase as shown in Fig. 4(e). Figure 4(e) is the amplitude sum $E_A$ and $E_B$ as a function of $\varphi$, demonstrating the constructive interference at $\varphi = \pm\left(\frac{(2m+1)\pi}{2}\right)$. Unlike conventional cavity optics, the phase resolution enhancement in Fig. 4(e) is $10^3$ over the classical limit due to higher order PBWs as shown in Figs. 4(c) and (d). Figure 4(f) is the intensity $I_A$ for (e), where the intensity is enhanced by a factor of four as analyzed above: For details, see the Supplementary Information. Except for $\varphi = \pm\left(\frac{(2m+1)\pi}{2}\right)$, the output intensity $I_A$ becomes zero due to destructive interference (see Fig. 4(g)).

Figure 4(h) is for a bad cavity mirror $(r = 0.9)$, where the effective n is also reduced. As expected from Figs. 4(c)~(e), the phase resolution is severely decreased from $10^{-4}$ to $10^{-2}$ due to low r-induced less n. However, the intensity $I_A$ is close to the upper bound. Unlike the classical resolution $\delta f_C$ limited by $\lambda_0/4$ (π/2) on a Sagnac interferometer (see Fig. 2(b)), the present cavity PBSI of Fig. 4(f) results in $\delta f_{PBSI} = \frac{\delta f_C}{2n}$. Regarding $E_B$ of equation (6-2), the sine function-induced asymmetric feature (the red-dotted curve in Fig. 4(e)) results in double side bands in its intensity $I_B$ (the red-dotted curve in Fig. 4(f)). So does the intensity correlation $g^{(2)}$ between the outputs $I_A$ and $I_B$ (not shown; see the Supplementary Information).

## 5. DISCUSSION

The mechanism of PBW in the present research is unique and completely different from conventional ones, where the nonclassical light generation of PBW is based on wave nature using double path superposition in a CCD-MZI. Thus, a serial connection of CCD-MZI can results in higher order PBWs. The novelty of this paper is not only new physics of coherence PBW but also applying it for Sagnac interferometer, where the phase resolution of the output light is automatically enhanced via many-wave interference for all orders of PBWs at a specific phase condition. Thus, conventional quantum metrology limited by nonclassical light such as higher order entangled photons and Schrödinger's cat can be overcome and applied for such as quantum lithography and interferometry.

Over decades, development of high accuracy inertial navigation systems has been raced in the area of ring laser gyroscope [30-32] and atom interferometry [21,33,34]. As limited to the SI, the size of ring gyro varies from ~1 [32] to ~$10^3$ (UG-2) [33] in $m^2$ depending on its purpose. The ring cavity stability has been well progressed to keep the thermal expansion bellow $10^{-8}$ K$^{-1}$ resulting in the random walk error of n°/$\sqrt{h}$ [31]. Such a high stability in a larger ring gyro can be compared with its small counterpart such as Honeywell GG 1839 whose stability is 200 μ°/$\sqrt{h}$.

On the contrary, atom interferometer Sagnac gyroscope has demonstrated earth's rotation rate sensing in the order of 30 ppm for the absolute geodetic rotation measurement [34]. The importance of earth rotating sensing is in geodesy and inertial navigation to detect such as Chandler wobble causing polar motion by unstable Earth rotation at very low frequency (26 nHz). Currently the sensitivity of atom interferometry is ~$10^{-9}$ rad/$\sqrt{s}$ [21]. Using the G ring whose cavity quality factor is $10^{12}$ [32], the theoretical estimation of sensitivity is $\Delta\Omega/\Omega_E \leq 10^{-8}$, where $\Delta\Omega$ is the quantum noise on resolution and $\Omega_E$ is the Earth rotation rate. Because the cavity in PBSI is basically the same as any ring cavity gyros, the presented PBSI can take over the state of the art in ring gyro systems for geodesy in a compact and portable unit.

## 6. CONCLUSION

In conclusion, the coherence version of PBW in a CCD-MZI was presented for fundamental physics and its potential applications to quantum metrology such as a cavity PBSI. The coherence PBW was analyzed and compared with the conventional quantum PBW, whose phase resolution enhancement is due to the ordered quantum superposition in a recursive CCD-MZI configuration. Such a recursive configuration was achieved in a cavity Sagnac interferometer whose antiphase condition in CCD-MZI is automatically satisfied by the Sagnac effect. Therefore, this work intrigues both communities of quantum physics and Sagnac interferometer to think about the origin of nonclassicality as well as implementation of quantum metrology without entangled photons. The present coherently driven nonclassical feature of PBW contributed to the enhanced phase resolution due to constructive interference among ordered components of PBW, where the *Finesse*-determined effective order n plays a key role. Although the enhanced phase resolution looks similar to conventional Febry-Perot type interferometer, the physics of PBSI is originated in the many-wave interference of PBW. The design of the cavity PBSI is pretty simple but smart to offer unprecedented phase resolution far beyond the classical limit of Rayleigh criterion. The cavity PBSI may open a door to a new realm of coherence-quantum metrology in the fields of gyroscope, inertial navigation, lithography, and geodesy. Owing to the huge enhancement factor in the phase resolution, the cavity PBSI can also be applied for nanophotonic optical gyro platforms [35] applicable to drones and robots with a stand-alone inertial system. Even for quantum lithography, the enhanced phase resolution in PBSI should be applicable to the semiconductor foundry.

References
1. W. M. Itano et al., "Quantum projection noise: Population fluctuations in two-level systems," Phys. Rev. A **47**, 3554-3570(1993).
2. V. Giovannetti, S. Lloyd, and L. Maccone, "Quantum-enhanced measurements: Beating the standard quantum limit," Science **306**, 1330-1336 (2004).
3. N. Kura and M. Ueda, "Standard quantum limit and Heisenberg limit in function estimation," Phys. Rev. Lett. **124**, 010507 (2020).
4. V. Giovannetti, S. Lloyd, and L. Maccone, "Quantum metrology," Phys. Rev. Lett. **96**, 010401 (2006).
5. L. Pezze, *et al.*, "Quantum metrology with nonclassical states of atomic ensemble," Rev. Mod. Phys. **90**, 035005 (2018).
6. J.-P. Wolf and Y. Silberberg, "Spooky spectroscopy," Nature Photon. **10**, 77-79 (2016).
7. O. Hosten, N. J. Engelsen, R. Krishnakumar, and M. A. Kasevich, "Measurement noise 100 times lower than the quantum-projection limit using entangled atoms," Nature **529**, 505-508 (2016)
8. K. J. Resch *et al.*, "Time-resolved and super-resolving phase measurements," Phys. Rev. Lett. **98**, 223601 (2007).
9. M. Xiao, L.-A. Wu, and H. J. Kimble, "Precision measurement beyond the shot-noise limit," Phys. Rev. Lett. **59**, 278-281 (1987).
10. M. J. Holland and K. Burnett, "Interferometric detection of optical phase shifts at the Heisenberg limit," Phys. Rev. Lett. **71**, 1355-1358 (1993).
11. J. Jacobson, G. Gjörk, I. Chung, and Y. Yamamato, "Photonic de Broglie waves," Phys. Rev. Lett. **74**, 4835-4838 (1995).
12. P. Walther, J.-W. Pan, M. Aspelmeyer, R. Ursin, S. Gasparon, and A. Zeillinger, "De Broglie wavelength of a non-local four-photon state," Nature **429**, 158-161 (2004).
13. X.-L. Wang, *et al.* "18-qubit entanglement with six photons' three degree of freedom," Phys. Rev. Lett. **120**, 260502 (2018).
14. S. F. Huelga, C. Macchiavello, T. Pellizzari, and A. K. Ekert, "Improvement of frequency standards with quantum entanglement," Phys. Rev. Lett. **79**, 3865-3868 (1997).
15. N. Samantaray, I. Ruo-Berchera, A. Meda, and M. Genovese, "Realization of the first sub-shot-noise wide field microscope," Light: Science & Applications **6**, e17005 (2017).
16. M. Kira, S. W. Koch, R. P. Smith, A. E. Hunter, and S. T. Cundiff, "Quantum spectroscopy with Schrodinger-cat states," Nature Phys. **7**, 799-804 (2011).
17. A. N. Boto, P. Kok, D. S. Abrams, S. L. Braunstein, "Quantum interferometric optical lithography: exploring entanglement to beat the diffraction limit," Phys. Rev. Lett. **85**, 2733-2736 (2000).
18. G. Sagnac, C. R. Acad. Sci. **157**, 708-710 (1913).
19. M. S. Shahriar, G. S. Pati, R. Tripathi, V. Gopal, M. Messall, and K. Salit, "Ultrahigh enhancement in absolute and relative rotation sensing using fast and slow light," Phys. Rev. A **75**, 053807 (2007).

20. B. Barrett *et al*. "The Sagnac effect: 20 years of development in matter-wave interferometry," C. R. Physique **15**, 875-883 (2014).
21. T. L. Bustavson, A. Landragin, and M. A. Kasevich, "Rotation sensing with a dual atom-interferometer Sagnac gyroscope," Class. Quantum Grav. **17**, 2385-2398 (2000).
22. K.-X. Sun, M. M. Fejer, E. Gustafson, and R. L. Byer, "Sagnac interferometer for gravitational-wave detection," Phys. Rev. Lett. **76**, 3053-3056 (1996).
23. E. J. Post, "Sagnac effect," Rev. Mod. Phys. **39**, 475 (1967).
24. B. S. Ham, "Deterministic control of photonic de Broglie waves using coherence optics: Coherence de Broglie waves," arXiv:2001.06913v4 (2020).
25. P. Grangier, G. Roger, and A. Aspect, "Experimental evidence for a photon anticorrelation effect on a beam splitter: A new light on single-photon interference," *Europhys. Lett.* **1**, 173-179 (1986).
26. B. S. Ham, "The origin of anticorrelation for photon bunching on a beam splitter," arXiv:1911.07174v2 (2019) (To be published in Sci. Rep.).
27. L. Mandel, "Coherence and indistinguishability," *Opt. Lett.* **16**, 1882-1883 (1991).
28. E. Solano, R. L. de Matos Filho, and N. Zagury, "Deterministic Bell states and measurement of the motional state of two trapped ions," *Phys. Rev. A* **59**, R2539-R2543 (1999)
29. A. Fried, M. Fejer, and A. Kapitulnik, "A scanning, all-fiber Sagnac interferometer for high resolution magneto-optic measurement at 820 nm," Review of Sci. Instument. **85**, 103707 (2014)
30. W. W. Chow, *et al*. "The ring laser gyro," Rev. Mod. Phys. **57**, 61-104 (1985).
31. K. U. Schreiber, et al., "How to detect the Chandler and the annual wobble of the Earth with a large ring laser gyroscope," Pure Appl. Geophys. **166**, 1485 (2009)
32. R. B. Hurst *et al.*, "Experiments with a 834 $m^2$ ring laser interferometer," J. Appl. Phys. **105**, 113115 (2009).
33. R. Geiger, *et al.* "Detecting inertial effects with airborne matter-wave interferometry," Nat. Communi. **2**, 474 (2011).
34. J. K. Stockton, K. Takase, and M. A. Kasevich, "Absolute geodetic rotation measurement using atom interferometry," Phys. Rev. Lett. **107**, 133001 (2011)
35. P. P. Khial, A. D. White, and A. Hajimiri, "Nanophotonic optical gyroscope with reciprocal sensitivity enhancement," Nature Photon. **12**, 671-675 (2018).